\documentclass[12pt,a4paper]{article}
\usepackage {amsmath}
\usepackage[dvips]{graphicx}
\usepackage{amssymb}
\usepackage{cite}
\newcommand{\be}{\begin{equation}}
\newcommand{\ee}{\end{equation}}
\newcommand{\bear}{\begin{eqnarray}}
\newcommand{\eear}{\end{eqnarray}}
\newcommand{\tanb}{\tan \beta}
\newcommand{\tanbb}{\tan^2 \beta}

\newcommand{\cosw}{\cos^2\theta_W}

\newcommand{\musq}[1]{ {\mu }^2 _{#1}}
\newcommand{\mhsq}[1]{ {m_{H _{#1}}^2}}

\newcommand{\tb}{\textbf}

\newcommand{\vev}[1]{\left\langle #1\right\rangle}

\newcommand{\lapproxeq}{\lower .7ex\hbox{$\;\stackrel{\textstyle  <}{\sim}\;$}} 
\newcommand{\gapproxeq}{\lower .7ex\hbox{$\;\stackrel{\textstyle  >}{\sim}\;$}} 
\newcommand{\stackdown}[2]{\lower 1.4ex\hbox{$\;\stackrel{\textstyle{#1}}  
{\scriptstyle{#2}}\;$}}
\newcommand{\bea}{\begin{eqnarray}} 
\newcommand{\eea}{\end{eqnarray}}

\makeatletter 
\def\slash{\@ifnextchar[{\fmsl@sh}{\fmsl@sh[0mu]}} 
\def\fmsl@sh[#1]#2{% 
  \mathchoice 
    {\@fmsl@sh\displaystyle{#1}{#2}}% 
    {\@fmsl@sh\textstyle{#1}{#2}}% 
    {\@fmsl@sh\scriptstyle{#1}{#2}}% 
    {\@fmsl@sh\scriptscriptstyle{#1}{#2}}} 
\def\@fmsl@sh#1#2#3{\m@th\ooalign{$\hfil#1\mkern#2/\hfil$\crcr$#1#3$}} 
\makeatother 
\begin{document} 
\allowdisplaybreaks

\begin{titlepage}  
%%%%%%%%%%% 
\begin{flushright} 
\parbox{4.6cm}{UA-NPPS/BSM-05/01 }
\end{flushright} 
%%%%%%%%%% 
\vspace*{5mm} 
\begin{center} 
{\large{\textbf {
Refinements in Effective Potential Calculations in the MSSM
}}}\\
\vspace{14mm} 
{\bf M. ~\ Argyrou}, \, {\bf A. ~\ Katsikatsou}  \, and 
 \\  {\bf I. ~\ Malamos} 

\vspace*{6mm} 
  {\it University of Athens, Physics Department,  
Nuclear and Particle Physics Section,\\  
GR--15771  Athens, Greece}

\end{center} 
\vspace*{15mm} 
%%%%%%%%%%%%%%%%%%%%%%%%% 
\begin{abstract}
The one-loop effective potential is a powerful tool in studying the 
electroweak symmetry breaking of supersymmetric theories, whose precise calculation may have
important phenomenological consequences. 
In this work, we are correctly treating the contribution of the Higgs sector to the effective potential and refine the radiative corrections to the Higgs mixing parameter $\mu$,
which is known to affect greatly the supersymmetric spectrum. 
Working at the average stop scale to minimize the effect of the stop sector, we find additional
corrections which can play a dominant role 
in the Focus Point region of the parameter space of the MSSM. 
The comparison of our results with those of the literature is discussed.
We also discuss the gauge dependence of the effective potential and its effect on the 
$\mu$ parameter in analyses where this is determined from the 1-loop minimization conditions of the
effective potential.
 
\end{abstract} 

\vspace{1cm}

\hspace{1cm}\hrule width 15 cm

\hspace{.2cm} {E-mails\,:\, margyrou@phys.uoa.gr, kkatsik@phys.uoa.gr, ymalam@phys.uoa.gr}
\end{titlepage} 
\newpage 
\baselineskip=20pt 
%%%%%%%%%%%%%%%%%%%%%%%%%% Paper body %%%%%%%%%%%%%%%%%%%%%%%%%
%%%%%%%%%%%%%%%%%%%%%%%%%%%%%%%%%%%%%%%%%%%%%%%%%%%%%%%%%%%%%%%

\section{Introduction}

\par The one loop effective potential \cite{Coleman:1973jx} is a powerful means in studying supersymmetric 
theories in order to extract information concerning the parameters describing the theory and consequently physical quantities,
e.g. the mass spectrum of the particles involved. Extensive studies have been done on this subject, and the radiative corrections to 
the Higgs masses 
have been computed using the effective potential approach and have been compared to those arising from the location of the poles of the propagators
\cite{Gamberini:1989jw,Haber,Ellis1,Barbieri,Brignole,Drees,deCarlos:1993yy,Dabelstein:1994hb,Chankowski:1992er} 
. The main purpose of this note is the proper treatment of the Higgs contributions to the one loop corrected potential, 
which although small may play a crucial role in the aforementioned studies in particular regions of the parameter space describing the supersymmetric models.
In our study we have as a prototype  the constrained Minimal Supersymmetric Standard
Model (CMSSM), involving terms that break supersymmetry (SUSY) softly. By
imposing universal boundary conditions at the unification scale $M_{GUT}$,
the model is defined by five parameters, namely, the common mass of all scalar fields
$m_0$, the common gaugino mass $M_{1/2}$ and the soft trilinear coupling $A_0$,
the ratio of the Higgs v.e.v 's, $\tanb$, and the sign of the Higgs mixing parameter $\mu$. 
The Higgsino and Higgs  mixing parameters $\mu, m_3^2$ are not free but they are determined through the minimization conditions. This is the line followed in numerous phenomenological analyses carried out so far. 
Our findings can be easily extended in cases where non-minimal cases  are considered, such as  departure  from the universal boundary conditions and/or extension to  
include CP violating phases and so on.

\par The total one-loop effective potential, at a reference scale Q, is $V_1(Q) \equiv V_0(Q) + \Delta V_1(Q)$, with  $V_0$  the tree level scalar potential, 
\bea
V_0(Q)  &=& m_1^2 (Q) \, \bigl( |H_1^0|^2 \, + \, |H_1^-|^2 \bigr) \, + \, m_2^2 (Q) \, \bigl( |H_2^0|^2 \, + \,  |H_2^+|^2 \bigr)\nonumber \\
 &+&   \bigl[ \, m_3^2 (Q) \, \bigl( H_1^0 \, H_2^0 \, - \, H_1^- \, H_2^+ \bigr) \, + \,  h.c. \, \bigr]\\ 
&+&	\frac{g^2 + g'^2}{8} \, \bigl( |H_1^0|^2 \, + \, |H_1^-|^2 \, - \, |H_2^0|^2 \, - \, |H_2^+|^2 \bigr)^2
\, + \, \frac{g^2}{2} \, \bigl( H_1^- \, H_2^{0*} \, + \, H_1^0H_2^{+*} \bigr)^2 \, , \nonumber 
\eea
where 
%$m_3^2 \, = \, \mu \, B$ 
and
$m_{1,2}^2 \, = \, m_{H_{1,2}}^2 \, + \, \mu^2$ and $m_{H_{1,2}}$ are the soft Higgs masses. $\Delta V_1$ is
the one-loop correction to the effective potential given by
\be 
 \Delta V_1 = {1\over{64\pi^2}} \,
\sum_J (-1)^{2s_J} \, (2s_J+1)\, m^4_J \,\left( \ln {m^2_J\over {Q^2}} 
- {3\over {2}} \right) \, .
\ee
In this $m_J$ are field dependent masses and $s_J$ denotes the
spin of the $J$-particle. 
The minimization of the one-loop corrected
effective potential $V_1$ yields the following
conditions \cite{Arnowitt:1992qp}: 
\be   \label{eq;min}
\sin 2\beta = -{ {2 m_3^2}\over{\bar{m}_1^2+\bar{m}_2^2} } \,\,, \,\,\,
\frac{M_Z^2}{2} = { {\bar{m}_1^2 - \bar{m}_2^2 \tan^2\beta} \over
 {\tan^2\beta-1} } \, , 
\ee
where 
%%%%%%%%%%%%%%%%%%%%%%%%%%%%%%%%%%%%%%%%%
\footnote{In our notation: 
$v_i \equiv \vev{H^0_i}$, $v_1 \equiv \dfrac{v}{\sqrt{2}} \cos\beta$,
$v_2 \equiv \dfrac{v}{\sqrt{2}}\sin\beta$,
$M_W^2=g^2(v_1^2+v_2^2)/2=g^2v^2/4$ .}
%%%%%%%%%%%%%%%%%%%%%%%%%%%%%%%%%%%%%%%%%%
\bea
\bar{m}_i^2 &\equiv& m_i^2 + \Sigma_i \;, \;\;\;\;\;\;\;\; \tan {\beta} \, = \, \frac{\upsilon_2}{\upsilon_1} \; \;\;\;\;\;\ \mathrm{and} \nonumber\\ 
\Sigma_i &\equiv& \left. { {\partial V^1}\over{\partial (\mathrm{Re} H_i^0)^2 }} 
\right|_{\vev{H_i^0}} \,.
\eea

%%% Extra comments %%%%%%
\par From (\ref{eq;min}) it is found that the one loop corrected $\mu$ is related to its tree level expression by 
\be \label{eq;muloop}
\musq {loop} =\musq {tree}+ \frac{ {\Sigma_1 -\tanbb \;\; \Sigma_2}}{\tanbb -1}.
\ee
The value of $\musq {tree}$ is defined by the same 
minimization conditions with the loop corrections to the scalar potential set to zero, 
\be \label{eq;mutree}
\musq {tree}=-\frac{{M_Z}^2}{2}+\frac{\mhsq {1}- \tanbb \;\; \mhsq {2}}{\tanbb -1} .
\ee
However the loop corrections in eq. (\ref{eq;min}) are very important, at least in certain regions of the parameter space, and should be duly taken into account.

It should be stressed that the renormalization scheme we use throughout is the minimal subtraction  $\overline{\textrm{DR}}$ scheme and we consider the value of $\mu$
as an output stemming from the 1-loop minimization conditions of the effective potential, eq. (\ref{eq;min}). 
This is the procedure followed in many analyses found in the literature. 
When working in such a scheme the gauge independence of $\tanb$ is not guaranteed,
unlike  "on-shell" schemes (OS) where 
$\dfrac{\delta \upsilon_1 }{\upsilon_1 } \, = \, \dfrac{\delta \upsilon_2}{\upsilon_2}$ 
is imposed by adjusting properly the counterterms \cite{Dabelstein:1994hb,Chankowski:1992er}. 
In the $\overline{\textrm{DR}}$ minimal subtraction scheme, keeping $\tan\beta$ fixed makes other parameters derived from 
the minimization conditions, notably the $\mu$ parameter, to be gauge dependent. 
This is not a problem since physical quantities after all, such as superparticle masses, are gauge independent if they are computed as poles of the 1-loop loop corrected propagators. However, at the tree level approximation, used sometimes for simplification purposes, some masses as for instance those of the chargino and neutralino, depend sensitively on $\mu$ and there is nothing to cancel its gauge dependence if tree level approximation is employed.
This may be a problem unless the gauge dependence of $\mu$ is imperceptibly small. This issue 
will be also discussed in this note. From the discussion above it becomes obvious that 
our findings are relevant to those who employ the $\overline{\textrm{DR}}$ minimal subtraction scheme, in the sense we described, rather than to OS- schemes users.

\par This paper is organized as follows. 
In section \ref{Landau}, we improve the analysis concerning 
the contribution of the Higgses to the quantities $\Sigma_{1,2}$ in the Landau gauge. 
The refined $\Sigma \, 's$ obtained are different from those
usually found in the literature \cite{Arnowitt:1992qp} and moreover, 
we find that they are important in regions of the 
parameter space where $\mu$ is small $\sim \mathcal{O} (M_Z)$,
as is the case in the Focus Point region. 
We also discuss the relevant one-loop corrections to the effective potential in the popular  't Hooft gauge and compare the $\Sigma \, 's$ obtained in the two gauges, 't Hooft and Landau. 
We find that differences are imperceptibly small, resulting to practically gauge independent values for the parameter $\mu$. 
In section \ref{mu}, we demonstrate the impact of these considerations 
on the calculation of the one-loop radiative corrections to the $\mu$ parameter
which greatly affects the spectrum, notably the neutralino and chargino sector as discussed previously. 
Finally, in section \ref{concl}, we present our conclusions.

\section{Improvement to $\Sigma_{1,2}$} \label{Landau}

\par 
The contribution of each particle sector to $\Sigma_{1,2}$ in the Landau gauge 
has been extensively discussed in ref. \cite{Arnowitt:1992qp}. 
The relevant quantities $\Sigma_{1,2}$ that enter the minimization conditions are defined 
through $\Sigma_i \, = \, \dfrac{1}{2 \, \upsilon_i} \; \biggl. \dfrac{\partial \Delta V}{\partial Re H_i^0}\, 
\biggr{|}_{\upsilon_i} \;, \; i=1,2$. 
%%%%%%%%%%%%%%
In using this, one should keep in mind that the substitution of the Higgs v.e.v's in this expression 
should be carried out after taking the derivatives of the 1-loop contribution 
to the effective potential $\Delta \, V$ 
with respect to the real parts of the Higgs fields, and not in the reverse order. 
Reversing the order affects the corrections stemming from the Higgs sector leaving those of the 
other sectors unaffected.
This, we think, has been overlooked in the literature and it is for that reason 
that our results concerning the Higgs bosons contributions 
differ from those quoted in ref. \cite{Arnowitt:1992qp}. 

Bearing this in mind, the contributions arising from the Higgs sector are found to be 
%%%%%%%%%%%%%%%%%%%%%%%%%%%
\bear
\Sigma _{{\rm{1 \,\,\, H}}{\rm{\,,\, h}}}^{Landau} \, &=& \, 
\frac{{\alpha _2 }}{{64\pi \cosw}}f\left( {M_{H \, , \, h}^2 } \right)\left( 1 \pm 2\frac{{\cos ^2 \beta M_Z^2  + (\sin ^2 \beta  - \cos 2\beta )M_A^2 }}{{M_H^2  - M_h^2 }} \right)
\nonumber\\
\Sigma _{{\rm{2\,\,\,   H}}{\rm{\,,\, h}}}^{Landau}  &=& \Sigma _{{\rm{1 \,\,\, H}}{\rm{\,,\, h}}}^{Landau} \, (\cos\beta \, \rightleftharpoons \, \sin\beta)
\eear
%%%%%%%%%%%%%%%%%%
where $M_{H, \, h}$ are the tree-level  masses of the heavy/light CP - even neutral 
Higgses respectively and $M_A$ that of the CP-odd Higgs given by
%%%%%%%%%%%%
\bear
M_{H, \, h}^2 \,& =& \, \frac{1}{2} \, \left[ (\, M_Z^2 \, + \, M_A^2 \,) \, \pm \, 
\sqrt{ (\, M_Z^2 \, + \, M_A^2 \,)^2 \, - \, 4 \,M_Z^2 \, M_A^2 \, cos^2 2 \beta} \, 
\right] \; \;  \\
M_A^2 \,&=&\, - \frac{2\, m_3^2}{sin 2 \beta} \, \, .
\eear  

As far as the contributions of the remaining Higgses are concerned, we have found that the CP-odd 
Higgs contributions are non-vanishing, contrary to what it is claimed in the literature, given by  
%%%%%%%%%%%%%%
\be
\Sigma_{1, A}^{Landau} \,= \,-\frac{{\alpha_2}}{64 \pi \cosw}\cos{2\beta}\,
f\left( {M_A^2 } \right), \qquad
\Sigma_{2, A}^{Landau}\, = \Sigma_{1, A}^{Landau} \, (\cos\beta \, \rightleftharpoons \, \sin\beta) \, , 
\ee
%%%%%%%%
while the corresponding charged Higgs contributions are given by 
%%%%%%%%%%%%%
\bear
\Sigma_{1, H_{\pm}}^{Landau}\,&=& \, \frac{{\alpha_2}}{32 \, \pi \, \cosw} \, ( \, 2 \, \cosw \, - \, \cos{2\beta}\, )
f\left( {M_{H_\pm}^2 } \right) \nonumber\\
\Sigma_{2, H_{\pm}}^{Landau}\,&=& \, \Sigma_{1, H_{\pm}}^{Landau}\, (\cos\beta \, \rightleftharpoons \, \sin\beta) \, \, .
\eear
In the equation above $M_{H_\pm} $ is 
the tree-level charged Higgs boson mass, $M_{H_\pm}^2 \, = \, M_A^2 \, + \, M_W^2$.
%%%%%%%%

%%%%%%%%%%%
In the expressions above the function $f \, ( \, m^2 \, )$ is defined by
%%%%%%%%%%%
\be
f \, ( \, m^2 \, ) \, = \, 2 \, m^2 \, (ln\frac{m^2}{Q^2} \, - \,  1 )\,.
\ee 
The scale $Q$ we choose to be of the order of the geometric average of the stop masses, 
$Q \, \sim \, \sqrt{m_{\widetilde{t}_1} \, m_{\widetilde{t}_2}}$, as it is customary in the literature,
so as to minimize the $1$-loop stop contributions. Then we can focus only on the contributions from the Higgs particles and the vector bosons.

\par The quantities $\Sigma_{1,2}$ given above differ from those cited in the literature. 
In particular the contribution of the pseudoscalar Higgs $(A)$, is not vanishing as it is stated
in \cite{Arnowitt:1992qp}. The amount of their difference will be discussed 
and quantified in the following. As far as the gauge bosons contribution is concerned, 
our findings are in agreement with those mentioned in the above reference.

Concerning the gauge dependence of $\Sigma_{1,2}$ we shall consider the one-loop 
effective potential in the t' Hooft gauge and compare it with the corresponding Landau gauge results.
The t' Hooft gauge is the one employed in many analyses and this is the reason it is chosen for a comparison
against the Landau gauge.
Lacking a direct calculation of the effective potential in the 't Hooft gauge we shall rely 
on the tadpole calculation to evaluate $\Sigma_{1,2}$ in this gauge. The relation between $\Sigma_{1,2}$ and the one loop tadpole graphs, in the same gauge which we choose to be 't Hooft's gauge, is given by 
%%%%%%%%%%%
\bear
\Sigma_i^{Hooft}
\, = \, - \, \frac{t_i}{\upsilon_i} \quad .
\; , \;\;\;\;\; i=1,2
\eear
%%%%%%%%%%%%
In this  $\upsilon_i$ are the v.e.v's and $t_i$ are defined to be the 1-loop tadpoles divided 
by the vertex factor $i \, (2\pi)^D \, \mu^{- 2 \varepsilon}$, where $\varepsilon=2-D/2$.
The choice of the gauge affects the $\Sigma_i \, 's$ only through the Higgs and gauge boson
contributions. We define the differences $\delta_i$ between Landau and 't Hooft gauge results by
%%%%%%%%%%
\be \label{eq;LanHoo}
\Sigma_i^{Landau} \; = \; \Sigma_i^{Hooft} \; + \; \delta_i \, \, .
\ee
%%%%%%%%%%%%%
That is $\delta_i$ denotes the amount of difference that arises in passing 
from one gauge to the other. In the Appendix, we present the explicit expressions for the  $\delta_i$'s stemming from the gauge boson and Higgs particles contribution at 1-loop order separately for each Higgs species in the $\overline{DR}$ scheme. 
In the following section we shall discuss the impact of the differences found on the parameter $\mu$ and quantify our results. 
%\newpage

%%%%%%%%%%%%%%%%%%%%%%%%%%%%%%%%%
%%%%%%%%%%%%%%%%%%%%%%%%%%%%%%%%%
\section{The $\mathrm{\mu}$ parameter at 1-loop} \label{mu}

The Higgs mixing parameter plays a vital role for the phenomenology of the MSSM, 
affecting particularly the neutralino and chargino sector. 
By defining
$ {\Delta \mu}\equiv \sqrt{\musq {loop}- \musq {tree}} \,$, 
one can estimate the size of the $1$-loop effects. In the following we actually study 
the dimensionless  ratio $\dfrac{{\Delta \mu }}{\mu_{tree} }$ as  
a function of the pseudoscalar mass $M_A$. $ {\Delta \mu}$ 
is influenced, in general, by both differences in $\Sigma_{1,2}$
and gauge differences as discussed in section \ref{Landau}.
In the Landau gauge,
\be
{\Delta \musq {}} \, = \, \frac{ {\Sigma_1 -\tanbb \;\; \Sigma_2}}{\tanbb -1} \, 
\ee
where $\Sigma_i \equiv \Sigma_i^{Landau}$ are  as given in section \ref{Landau}. In the same gauge, according to \cite{Arnowitt:1992qp}, the same quantity is
\be
({\Delta \mu^2})' = \frac{ {\Sigma_1' -\tanbb \;\; \Sigma_2'}}{\tanbb -1} \,.
\ee
where primes  denote the contributions to the $1$-loop effective potential, 
as given in the appendix of reference\cite{Arnowitt:1992qp}.

Concerning the gauge differences, the relation between the $\Delta \mu $ 's defined above are given by, in an obvious notation,
\be
{\Delta \musq {Hooft}}={\Delta \musq {Landau}}-\frac{{\delta_1 -\tanbb \;\; \delta_2 }}{\tanbb -1} \,.
\ee
where $\delta_i$ are those of eq.(\ref{eq;LanHoo}) which are analytically given in the Appendix.

In order to quantify these differences one needs the soft Higgs mass parameters $\mhsq {1,2}$, which define $\musq{tree}$, 
at the electroweak scale. For their evolution one can use the findings of ref. \cite{Ibanez:1984vq}. 
This is known to be valid for low $\tan\beta$ and it can be used if one wants to derive 
analytic expressions in exploring particular regions of the parameter space. 
In this $\tanb$ regime, one can safely neglect the evolution of Yukawa couplings 
for the bottom quark and the tau lepton and the following 1-loop expressions hold :
\bear
\mhsq {1} (t)&=& m_0^2+{M_{1/2}}^2 g(t) \\
\mhsq {2} (t)&=& {M_{1/2}}^2\,e(t)+ A_0\,M_{1/2}\, f(t)+ \left( m_0^2 \,h(t)- A_0^2 \,k(t) \,\right)
\eear
The variable $t$ parametrizes the energy scale, $ t = ln\dfrac{M_{GUT}^2}{M_Z^2} $
and the functions $g,e,f,h,k$ are given in the above reference. We are aware of the fact 
that this approximation holds for low values of $\tan\beta$ as stated previously. 
However, these analytic expressions help us locate the Focus Point region \cite{Nath97}, the significance of which
is emphasized in the following. For larger values of $\tanb$, the b and $\tau$ contributions 
are important and should not be neglected. 
In our numerical approach, for larger values of $\tanb \gtrsim 7$, these 
contributions are duly taken into account. In fact we numerically solve the $2$-loop RGEs
of the parameters involved and we also take into consideration other SUSY
corrections to the Yukawa couplings which start being important for large values 
$\gtrsim 40$ of $\tanb$.
Then, the tree level mass of the pseudoscalar Higgs (A), entering the expressions for 
$\Sigma's$, at the electroweak scale
is given by
\be
M_A^2 \, = \, \mhsq{1} \, + \, \mhsq{2} \, + \, 2 \, \musq{tree} .
\ee

In order to demonstrate the importance of the differences mentioned before, 
we focus on these regions of the parameter space where the value of the ratio 
$\dfrac{{\Delta \mu }}{\mu_{tree} }$ becomes substantial. 
It is in this region where the effects are expected to be enhanced. 
This occurs when $\mhsq{i} \, \sim \, \Sigma_i$. Therefore, we restrict 
our analysis to these domains of the parameter space which are characterized by rather 
large values of $\tanb \gtrsim 7$ and referred to as the hyperbolic branch (HB) 
of the radiative breaking \cite{Nath97}. 
In this region, the loop corrections to $\mu$ are significant. 
The soft parameters, $m_0$ and/or $M_{1/2}$ can take very large values while 
$\mu$ stays relatively small, of the order of the electroweak scale. 
This last remark is of a particular interest, since we have estimated 
that $(\Delta \mu^2)' \, - \, \Delta \mu^2 \sim \mathcal{O} (M_Z^2)$. 
Thus the differences found can be important only if $\mu$ 
is of the order of $M_Z$, as it occurs in a subset of the HB, 
the so called Focus Point (FP) region \cite{Matchev1, Matchev2}. 
This is characterized by relatively low values of $M_{1/2}$, 
large $m_0$, of the order of a few TeV  and the values for $\mu$ are close 
to the electroweak scale. 
Thus, in the following, we further limit our analysis to this particular region 
which is both phenomenologically and cosmologically interesting. Regarding the latter 
the LSP neutralino in this region is a mixture of Bino and Higgsino 
and the Higgsino impurity allows for rapid s-channel LSP annihilations, 
resulting to low neutralino relic densities, at experimentally acceptable levels.

\begin{figure}[h!]  
\begin{center}
\includegraphics[scale=.7]{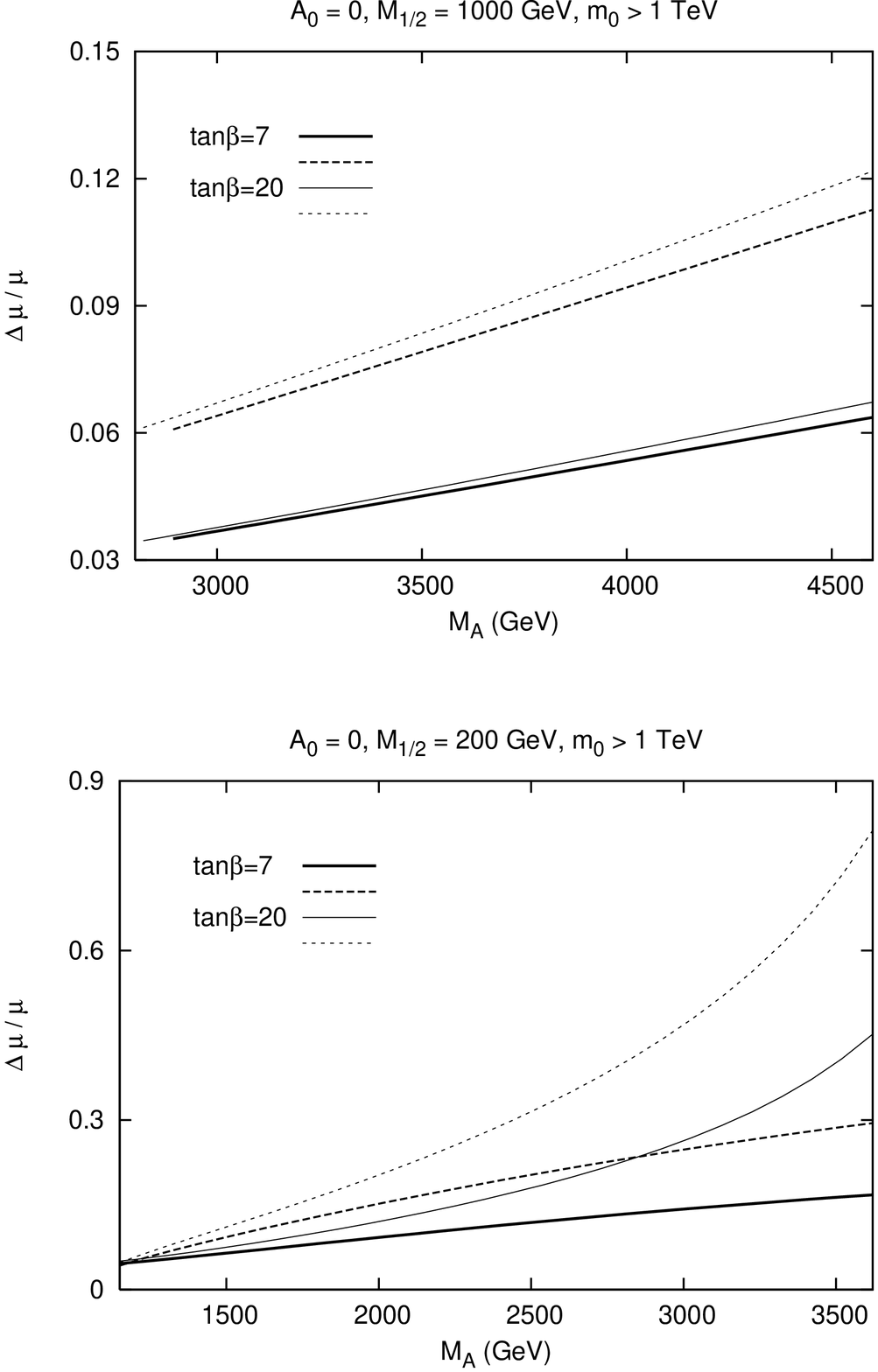} %\hspace{1cm}
\caption[]{\small The ratio ${\Delta \mu }/{\mu_{tree} }$ in the Landau gauge as 
           a function of $M_A$ for the inputs displayed in the figures and $\mu >0$. 
           The solid lines are the ratios ${\Delta \mu }/{\mu_{tree} }$ 
           according to section \ref{Landau}. 
           The dashed lines are the same ratios calculated 
           as in reference \cite{Arnowitt:1992qp}, denoted by 
           ${\Delta \mu ' }/{\mu_{tree}} \,$ in the main text.}
\label{fig1}  
\end{center}
\end{figure}

We first compare the ratios 
${\Delta \mu }/{\mu_{tree} }$, as calculated in this work, with   
${(\Delta \mu)' }/{\mu_{tree} }$, found in \cite{Arnowitt:1992qp}, both with the effective potential contributions calculated in the Landau gauge.
In fig.\ref{fig1}, we plot these ratios for two different input values 
of $\tan\beta$ and $M_{1/2}$. 
In every case, the solid lines are the ratios ${\Delta \mu }/{\mu_{tree} }$ and 
the dashed lines are the ratios ${(\Delta \mu)' }/{\mu_{tree} }$. 
In the upper panel of fig.\ref{fig1}, we plot these ratios as a function 
of the pseudoscalar mass $M_A$ for values of $\tan\beta = 7$ and $20$, 
respectively and for fixed $A_0 = 0$, and $M_{1/2} = 1$ TeV. 
In the lower panel, the $M_{1/2}$ value has been decreased to $M_{1/2} = 200$ GeV. 
In both panels, the values of $M_A$ have been obtained by random values of the $m_0$ parameter, 
in the region of 1 TeV to about 3.5 TeV. Recall that we have in mind the constrained MSSM scenario
so that the values of $M_A$ are outputs. 
In the first case ($M_{1/2} = 1 $ TeV) the differences are small. Actually in this case 
we are far from the Focus Point region which shows up for $M_{1/2}<< m_0$ values.
However, in the second case ($M_{1/2} = 200$ GeV), we are well within the Focus Point region, 
since $M_{1/2}<< m_0$ and we observe that ${(\Delta \mu)' }/{\mu_{tree} }$ is greatly enhanced, 
reaching values as large as $80 \%$. 
Note, however, that the refined ratio ${\Delta \mu }/{\mu_{tree} }$, although still large, 
is almost half of it, never exceeding $45\%$. Thus such differences may affect the SUSY spectrum and, in particular, 
the neutralino and chargino sectors, and less the stop and sbottom mass spectrum, through the left/right 
squark mixings which depend on $\mu$. We have estimated the relative difference 
of the $1-$loop corrected masses of the chargini and the lightest of neutralini 
with respect to their tree-level values to be about $20 \%$ whereas according 
to the approach followed in \cite{Arnowitt:1992qp}, this difference can reach $30-40\%$.

Last, in order to illustrate the effect of the gauge differences between 
the Landau and the 't Hooft gauge on the $\mu$-parameter, 
we consider again the ratios ${\Delta \mu }/{\mu_{tree} }$ in the two gauges.
We find that the Higgs contribution is insensitive to the gauge choice, as shown 
in the upper panel of fig.\ref{fig2}. For the gauge bosons (lower panel), 
we observe a difference which is almost tripled, in passing from the 't Hooft 
to the Landau gauge, but the corresponding ratios ${\Delta \mu }/{\mu_{tree}}$ 
are small, at per mille level, unable to account for a sizable effect 
on $\mu$. Therefore gauge dependences of $\mu$, 
which is a gauge dependent quantity in approaches that it is not an input but it is derived from the minimization conditions,
are quite small. In view of this, tree level masses are almost gauge independent despite the fact that they may depend on $\mu$.
This justifies their use as approximations to physical masses whenever needed.

\section{Conclusions}
\label{concl}

We refine the calculation of the Higgs contributions to the effective potential in the constrained supersymmetric model with universal boundary conditions at the unification scale. Our results can be easily generalized to non-minimal models. 
We demonstrate the impact of this calculation on the one-loop corrected  $\mu$ parameter 
and its consequence on the mass spectrum, in particular on the chargino and neutralino 
masses in certain regions of the parameter space. 
The estimated results are shown to be quite different, as compared to those presented 
in the literature, when we are within the Focus Point region of the parameter space describing the CMSSM. 
We also discuss the effect of the gauge dependence on the $\mu$ parameter
as this is determined by the minimization conditions 
in the 't Hooft and the Landau gauges. 
We found that the differences between the two gauges are extremely small,
of the order of $1\%$, due mainly to the gauge bosons contribution.
Thus the $\mu$- dependent tree level masses are almost gauge independent
and can be used as physical masses if they are the dominant contributions
to the propagators.    \\[.8cm]
{\bf \large Acknowledgments}\\
We wish to thank A.B. Lahanas for enlightening discussions, thoroughly reading \\
the manuscript and continuous support during this effort. 
This work is co-funded by the European Social Fund $(75 \%)$ 
and National Resources $(25 \%)$-(EPEAEK)-PYTHAGORAS.
The work of A.K. was also supported by the Greek State Scholarship's Foundation (IKY).
%
%%%%% BEGIN APPENDIX%%%%%%%%%%%%
\appendix
\section*{Appendix} \label{app}

\textbf{CP-even Higgs H, h}

\par The amount of differences in the $1$-loop effective potential 
between Landau and 't Hooft gauge in the $\overline{DR}$ scheme, 
are found to be
%%%%%%
\be
\delta _i^{H,h}= \frac{{\alpha _2}}{{64\pi \cos ^2 \theta _W }}(f\left( {M_H^2 } \right)- 
f\left( {M_h^2 } \right)){\mathcal{A}}_i \;,\; \; \; i=1,2
\ee
%%%%%%%%%
where 
\bear
\mathcal{A}_1=2\,\frac{{\cos ^2 \beta \,M_Z^2+(\sin ^2 \beta  - \cos 2\beta )\,M_A^2 }}{{M_H^2  - M_h^2 }}- 2\cos 2a + \sin 2a\,\tan \beta \nonumber\\
\mathcal{A}_2=2\,\frac{{\sin ^2 \beta \,M_Z^2+(\cos ^2 \beta  + \cos 2\beta )\,M_A^2 }}{{M_H^2  - M_h^2 }}+ 2\cos 2a + \sin 2a\cot \beta 
\eear
The CP-even Higgs mixing angle $\alpha$ is given at the tree level by
%%%%%%%%%
\be
tan 2 \alpha \, = \, tan 2 \beta \, \frac{M_A^2 \, + \, M_Z^2}{M_A^2 \, - \, M_Z^2}
\ee
%%%%%%%%%%%%%
Loop corrections are known to have an important effect on $\alpha$ and 
this has been taken into account in our numerical procedure. 

\noindent
\tb{Charged Higgs}
\be
\delta_1^{H_{\pm}} \, = 0 \; , \;\;\;\;\;\;\;\;\;\;\; \delta_2^{H_{\pm}} \, = \, 0
\ee
\tb{CP-odd Higgs A}
\be
\delta_1^A \, = \, 0 \;\;\;\;\; , \;\;\;\;\;\ \delta_2^A \, = \, 0 
\ee
%%%%%%%%%%%%%%%
%
\tb{Gauge bosons}
\be
\delta_1^{gauge} \, = \, - \, \frac{\alpha_2 \, cos 2 \beta}{32 \, \pi \, cos^2 \theta_W} \,  [ \, f(M_W^2) \, + \, \frac{1}{2} \, f(M_Z^2) \, ] \; , \;\;\;\;\;\;\;\;\;\;\; \delta_2^{gauge} \, = \, - \, \delta_1^{gauge}
\ee
%%%%%%% END APPENDIX&&&&&&&&&&&&&

\newpage

\begin{figure}[h!]  
\begin{center}
\rotatebox{270}{\includegraphics[scale=.45]{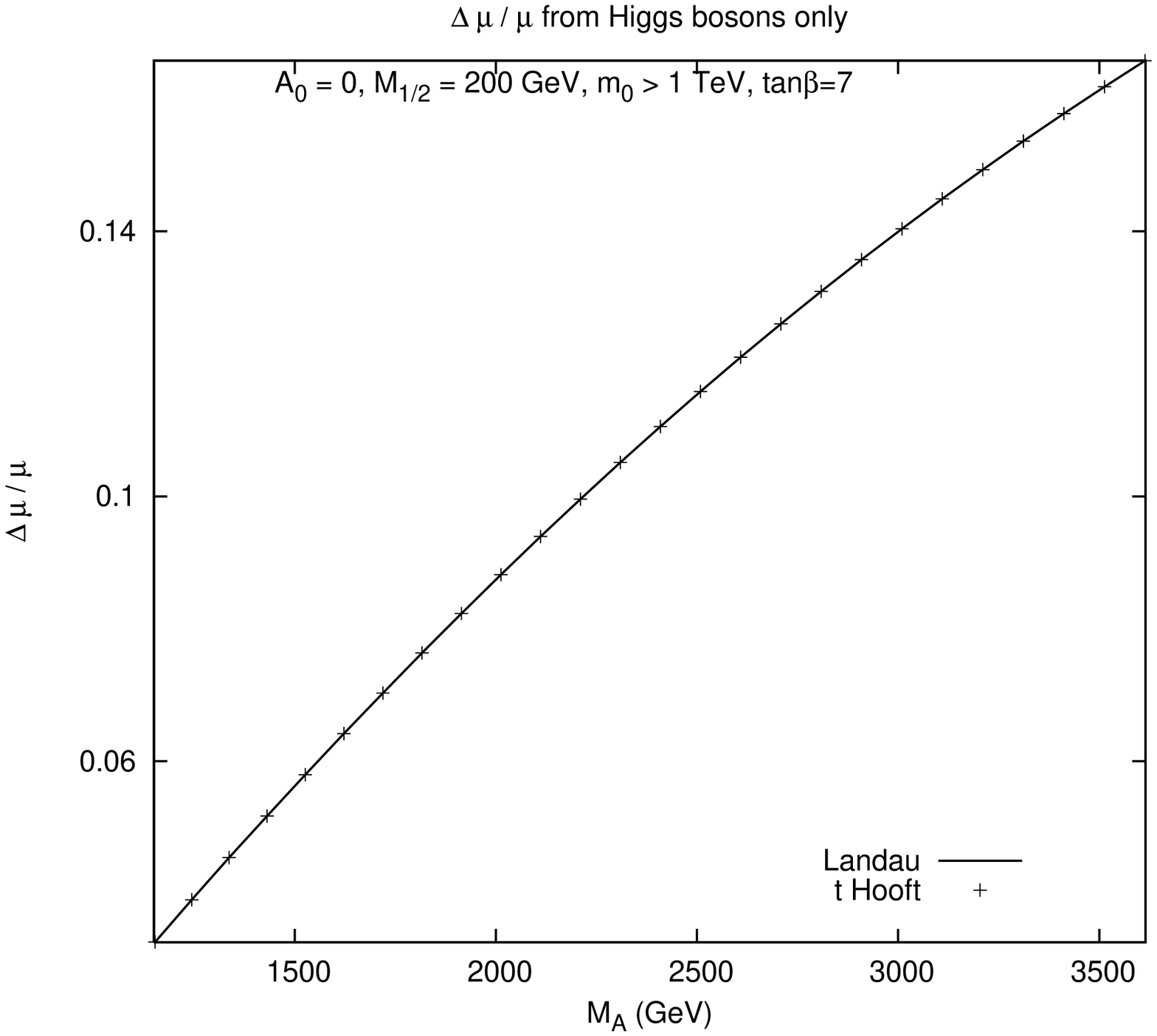}} %\hspace{1cm}
\end{center}
\end{figure}

\begin{figure}[h!]  
\begin{center}
\rotatebox{270}{\includegraphics[scale=.45]{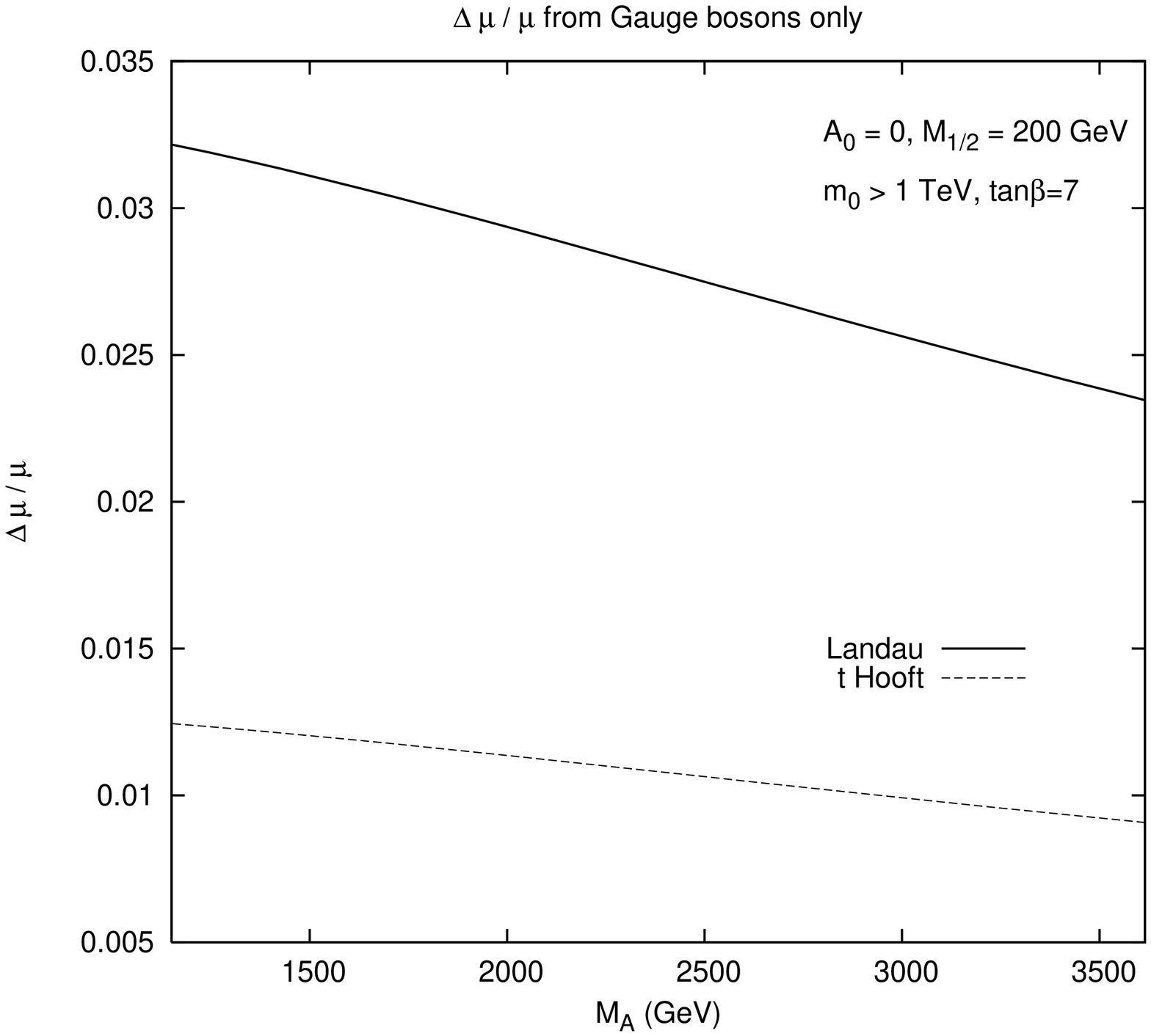}} %\hspace{1cm}
\caption[]{\small The ratio ${\Delta \mu }/{\mu_{tree} }$ calculated both in the Landau 
and in the 't Hooft gauge as a function of $M_A$ for the inputs displayed in the figures and $\mu >0$. 
In $\Delta \mu$, the contribution of the Higgs bosons (upper panel) and the vector bosons 
(lower panel) are separately displayed.  
The results in the Higgs case are almost identical as seen in the top panel.}
\label{fig2}  
\end{center}
\end{figure}

\end{document}